\documentstyle[prb,aps,twocolumn,epsf,floats]{revtex}
\begin{document}
\draft
\twocolumn[\hsize\textwidth\columnwidth\hsize\csname @twocolumnfalse\endcsname
\title{Re-examination of orbital ordering in LaMnO$_3$: \\ 
Electron-electron and electron-lattice interactions}
\author{Satoshi~Okamoto}
\address{The Institute of Physical and Chemical Research (RIKEN), Saitama 351-0198, Japan}
\author{Sumio~Ishihara}
\address{Department of Applied Physics, University of Tokyo, Tokyo 113-8656, Japan}
\author{Sadamichi~Maekawa}
\address{Institute for Materials Research, Tohoku University, Sendai 980-8577, Japan}
\date{\today}
\maketitle
\begin{abstract} 
The interactions between $e_g$ orbitals in neighboring sites are investigated in LaMnO$_3$ 
by taking into account virtual exchange of electrons and phonons. 
The spin and orbital ordering temperatures and the spin wave dispersion relation are calculated. 
We find that the orbital ordering is mainly caused by the electronic interactions and 
that the Jahn-Teller coupling is much smaller than that reported previously. 
We propose that the elastic constant shows a characteristic change at the N{\'e}el temperature 
by the spin and orbital coupling and the higher-order Jahn-Teller coupling.
\end{abstract}
]
\narrowtext
\noindent
\section{Introduction}
In some classes of transition metal oxides, 
degeneracy of the $d$ orbitals of a transition metal ion 
remains and electrons have a degree of freedom indicating the occupied orbital. 
This is called orbital degree of freedom.\cite{science} 
For the colossal magnetoresistance (CMR) 
observed in perovskite manganese oxides,\cite{chahara,helmolt,tokura} 
the orbital degree of freedom is considered to play an important role, 
because the gigantic decrease of resistivity is observed in the vicinity of the transition 
from charge and orbital ordered phase to ferromagnetic metallic one. 
A parent compound of the CMR manganites, LaMnO$_3$, shows the orbital ordering below 
780~K associated with the distortion of a MnO$_6$ octahedron. 
It has been experimentally confirmed that the type of the orbital ordering is of 
$C$-type\cite{murakami} where two kinds of orbitals are alternately aligned in 
the $xy$ plane and the planes are stacked along the $z$ axis. 
In addition to the orbital ordering, 
the so-called $A$-type antiferromagnetic (AF) ordering appears below 145~K, 
where spins are aligned parallel (antipararell) in the $xy$ plane 
($z$ axis).\cite{wollan,matsumoto}  
It is well recognized that this anisotropic magnetic ordering is 
stabilized by the orbital ordering.\cite{murakami,wollan,matsumoto,goodenough,kanamori,kugel,hirota,ishihara1,rodriguez,maezono,feinberg} 
\par
In $3d$ transition metal compounds with orbital degeneracy, 
two kinds of mechanisms have been proposed for the orbital ordering. 
One is caused by the superexchange (SE)-type interaction between orbitals 
in different sites. 
This interaction originates from the virtual exchange of electrons under 
the strong on-site electron-electron interactions.\cite{kugel,ishihara1} 
Another mechanism of the orbital ordering is based on the cooperative Jahn-Teller (JT) 
effects where the lattice distortion occurs cooperatively 
and lifts the orbital degeneracy in the transition metal ions.\cite{kanamori2,kataoka,millis,kataoka2}
The effective interaction between orbitals in this mechanism is caused by 
virtual exchange of phonons. 
However, it is usually difficult to separate contributions of these two mechanisms 
to the orbital ordering. 
This is because the two mechanisms provide the effective orbital interactions 
cooperatively.\cite{nagaosa,benedetti,bala} 
\par
It has been supposed that the strong electron-lattice 
interaction exists and is necessary to explain CMR. 
The orbital ordering in LaMnO$_3$ was also studied based on the cooperative JT effects.\cite{kanamori2,millis,kataoka2} 
The energy splitting of the two $e_g$ orbitals due to the lattice distortion termed 
the JT energy ($E_{JT}$) was estimated to be of the order of 1eV 
by analyzing the orbital ordering temperature,\cite{millis} 
optical spectra\cite{jung,machida,quijada,allen,coey,ahn} 
and the energy band calculation.\cite{satpathy,popovic,hozoi} 
We note that in these analyses, the electron correlation effect was not taken into account properly. 
Actually, the on-site Coulomb interaction between electrons was estimated experimentally 
to be about 7eV which is much larger than $E_{JT}$.\cite{t.saitoh} 
Therefore, it is necessary to re-examine the orbital ordering in LaMnO$_3$ 
by considering both the cooperative JT effect and the SE 
interaction under the strong electron correlation on an equal footing. 
\par
In this paper, we investigate the interactions between $e_g$ orbitals in neighboring sites 
originating from the electron-electron and electron-lattice interactions in LaMnO$_3$. 
Magnitudes of these interactions are determined through the calculation of 
the spin and orbital ordering temperatures and the spin stiffness constants. 
It is shown that $E_{JT}$ is much smaller than that 
in the literature\cite{millis,jung,machida,quijada,allen,coey,ahn,satpathy,popovic,hozoi} and 
the orbital ordering is mainly caused by the electronic interactions. 
We find that the elastic constant shows a characteristic change at the N{\'e}el temperature 
by which the coupling constant of the higher-order JT effect is estimated. 
\par
In Sec.~II, the model Hamiltonian which describes the orbital interactions 
caused by exchanges of electrons and phonons is derived. 
In Sec.~III, we introduce the mean field approximation in the formulation of 
the orbital and spin ordering temperatures. 
In Sec.~IV, by comparing the theoretical results of the ordering temperatures and 
the spin stiffness constants with the experimental values, 
the magnitudes of the orbital interactions are determined numerically. 
Temperature dependence of the elastic constants are studied in Sec.~V. 
The last section is devoted to the summary and discussion. 
\section{Model} 
We start with the following Hamiltonian which includes spin, orbital and lattice 
degrees of freedom: 
\begin{eqnarray}
{\cal H}= {\cal H}_{e} + {\cal H}_{e-latt} + {\cal H}_{latt} 
+ {\cal H}_{str} + {\cal H}_{e-str} + {\cal H}_{hiJT}.
\label{eq:Htot}
\end{eqnarray}
${\cal H}_e$ describes the electronic interactions and consists of three terms as 
\begin{eqnarray}
{\cal H}_e= {\cal H}_J + {\cal H}_H + {\cal H}_{AF}.
\label{eq:Heff}
\end{eqnarray}
${\cal H}_J$ represents the SE interaction 
between nearest neighboring (NN) $e_g$ electrons 
derived from the generalized Hubbard model with orbital degeneracy\cite{ishihara1} as 
\begin{eqnarray}
{\cal H}_J 
=\!\! &-&2J_1\sum_{\langle ij \rangle } 
 \biggl ( {3 \over 4} n_i n_j + \vec S_i \cdot \vec S_j \biggr )
 \biggl ( {1 \over 4}  - \tau_i^l \tau_j^l \biggr ) \nonumber \\
 \!\! &-&2J_2\sum_{\langle ij \rangle } 
 \biggl ( {1 \over 4} n_i n_j  - \vec S_i \cdot \vec S_j   \biggr ) 
 \biggl ( {3 \over 4}   + \tau_i^l \tau_j^l +\tau_i^l+\tau_j^l \biggr ), 
\label{eq:Hj}
\end{eqnarray}
where 
$J_1 = t_0^2/(U' - I)$ and $J_2 = t_0^2/(U' + I + 2J_H)$. 
$U, U'$ and $I$ are the intra-, inter-orbital Coulomb interactions and the exchange interaction 
for $e_g$ electrons, respectively, 
and a relation $U = U' + I$ is assumed. 
$J_H$ is the Hund coupling between $e_g$ electron and $t_{2g}$ spin $\vec S_i^t$ ($S^t=3/2$), 
and $t_0$ is the transfer intensity between NN $d_{3z^2-r^2}$ orbitals along the $z$ axis. 
Energy splitting between two $e_g$ orbitals due to the JT effect is neglected 
in the denominators of $J_1$ and $J_2$, 
because this splitting is much smaller 
than the Coulomb interactions.\cite{millis,jung,machida,quijada,allen,coey,ahn,satpathy,popovic,hozoi,t.saitoh} 
$\vec{S}_i$ is the spin operator of an $e_g$ electron with $S=1/2$. 
$\tau_i^l$ is defined as 
$\tau_i^l = \cos (\frac{2\pi}{3}m_l) T_{iz} - \sin (\frac{2\pi}{3}m_l) T_{ix}$ 
with $(m_x, m_y, m_z) = (1, -1, 0)$ where 
$l$ denotes a direction of a bond connecting sites $i$ and $j$. 
$\vec{T}_i$ is the pseudospin operator for the orbital degree of freedom, and 
$\langle T_{iz} \rangle= +(-)1/2$ corresponds to 
the state where the $d_{3z^2-r^2}$ ($d_{x^2-y^2}$) orbital is occupied by an electron. 
The second and third terms in Eq.~(\ref{eq:Heff}) describe 
the Hund coupling between $e_g$ and $t_{2g}$ spins 
and the AF SE interaction ($J_{AF}$) between NN $t_{2g}$ spins, respectively. 
These are given by 
\begin{eqnarray}
{\cal H}_{H} + {\cal H}_{AF}
= -J_H \sum_i \vec{S}_i \cdot \vec{S}^t_i 
+J_{AF} \sum_{\langle i j \rangle} \vec{S}^t_i \cdot \vec{S}^t_j . 
\label{eq:hhhaf}
\end{eqnarray}
%
%
%
%
%
\begin{figure}
\epsfxsize=0.8\columnwidth
\centerline{\epsffile{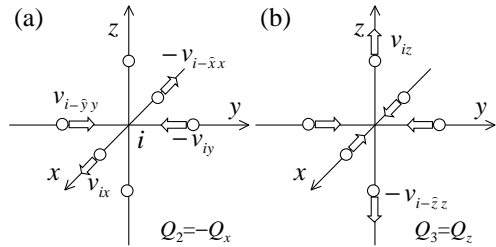}}
\caption{The modes of the distortion of a MnO$_6$ octahedron. 
(a) $Q_2$ and (b) $Q_3$ modes. 
}
\label{fig:fig1}
\end{figure}
%
%
%
%
\begin{figure}
\epsfxsize=0.75\columnwidth
\centerline{\epsffile{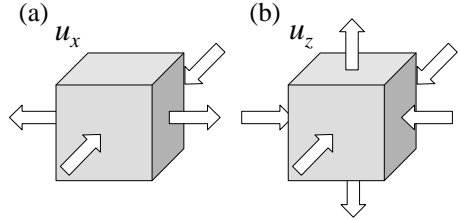}}
\caption{The modes of the bulk strain. 
(a) $u_x$ and (b) $u_z$ modes. 
}
\label{fig:fig2}
\end{figure}
%
%
The second and third terms in Eq.~(\ref{eq:Htot}) describe 
the electron-lattice interaction and the lattice dynamics, respectively. 
Here, we consider the displacement of O ions along the direction connecting NN Mn ions,  
since the motion of O ions along the other directions does not couple linearly 
with the $e_g$ orbitals. 
Thus, ${\cal H}_{e-latt}$ is given by 
\begin{eqnarray}
{\cal H}_{e-latt}=-g_{JT} \!\! \sum_{i\: l=x,z} \!\! Q_{il} T_{il}\, ,
\label{eq:Helatt}
\end{eqnarray}
where $g_{JT}$ is the coupling constant. 
$Q_{ix}$ and $Q_{iz}$ are the normal modes of the lattice distortion at site $i$ 
given by 
$Q_{ix}={1 \over \sqrt{2}} (-v_{ix}+v_{i-\hat x \, x}+v_{iy}-v_{i-\hat y \, y})$
and 
$Q_{iz}={1 \over \sqrt{6}} (2v_{iz}-2v_{i-\hat z \, z}-
v_{ix}+v_{i-\hat x \, x}-v_{iy}+v_{i-\hat y \, y})$. 
\cite{kanamori2}
$v_{i\xi}$ is the displacement of an O ion at $\vec r_i + {a \over 2}\hat \xi$ and 
$a$ is the lattice constant. 
These normal modes are schematically shown in Fig.~\ref{fig:fig1}. 
The third term in Eq.~(\ref{eq:Htot}), ${\cal H}_{latt}$, is given by 
\begin{eqnarray}
{\cal H}_{latt}=\sum_{\vec k \, \xi=x,y,z} {\hbar \omega_{\vec k} \over 2} 
\Bigl( p_{\vec k \xi}^*p_{\vec k \xi} + q_{\vec k \xi}^*q_{\vec k \xi} \Bigr), 
\label{eq:Hlatt}
\end{eqnarray}
where 
$q_{\vec k \xi}$ is the normal coordinate of lattice vibration 
with direction of displacement $\xi$ and momentum $\vec k$, 
and $p_{\vec k \xi}$ is the canonical conjugate momentum of $q_{\vec k \xi}$. 
$q_{\vec k \xi}$ and $v_{i\xi}$ satisfy the relation 
$v_{i\xi}={1 \over \sqrt{N}} \sum_i e^{i \vec k \vec r_i}
\sqrt{\hbar \over m \omega_{\vec k}}q_{\vec k \xi}$ 
with $N$ being the total number of Mn sites. 
The frequency of the lattice vibration is independent of $\vec k$ and is given by 
$\omega_{\vec k}=\sqrt{K \over m}$ with $m$ being the mass of an O ion, 
since only the spring constant $K/2$ between NN Mn and O ions is taken into account. 
%
${\cal H}_{str}$ and ${\cal H}_{e-str}$ in Eq.~(\ref{eq:Htot}) describe the elastic energy and 
electron-strain coupling, respectively,\cite{kataoka} as 
\begin{eqnarray}
{\cal H}_{str} = {V c_0 \over 2} \bigl( u_x^2 +u_z^2 \bigr), 
\label{eq:Hstr}
\end{eqnarray}
and 
\begin{eqnarray}
{\cal H}_{e-str}
= - 2 g_0 \sqrt{V c_0 \over N} \sum_i ( u_x T_{i x}+u_z T_{i z} ). 
\label{eq:He-str}
\end{eqnarray}
Here, $V$ is the volume of the system and 
$c_0$ is the elastic constant. 
The electron-strain coupling constant $g_0$ is related to $g_{JT}$ as 
$g_0 = {a \over 2} \sqrt{N \over V c_0} g_{JT}$. 
The bulk distortions $u_x$ and $u_z$ 
are represented by the elastic strain $e_{\rho \rho'} (\rho, \rho'=x,y,z)$ as 
$u_x={1 \over \sqrt{2}}(e_{yy}-e_{xx})$ and 
$u_z={1 \over \sqrt{6}}(2e_{zz}-e_{yy}-e_{xx})$, respectively.\cite{kataoka} 
Schematic pictures of the bulk distortions are presented in Fig.~\ref{fig:fig2}. 
%
The last term of Eq.~(\ref{eq:Htot}) describes the higher-order JT coupling given by 
\begin{eqnarray}
{\cal H}_{hiJT} =- B{V c_0 \over g_0^2 N} 
\sum_i \Bigl\{ \bigl( Q_{iz}^2-Q_{ix}^2 \bigr) T_{iz} - 2 Q_{iz}Q_{ix} T_{ix} \Bigr\},
\label{eq:Hhijt}
\end{eqnarray}
with coupling constant $B$. 
\par
Now, we derive the effective Hamiltonian describing the inter-site orbital interaction 
through the exchange of phonon\cite{kataoka} 
from ${\cal H}_{e-latt}$ and ${\cal H}_{latt}$. 
${\cal H}_{e-latt}$ is rewritten by using the Fourier transforms of 
$q_{\vec k\xi}={1 \over \sqrt{N}}\sum_i e^{-i \vec k \vec r_i} q_{i\xi}$ and 
$T_{\vec k l}={1 \over \sqrt{N}} \sum_i e^{-i \vec k \vec r_i} T_{i l}$ 
as 
\begin{eqnarray}
{\cal H}_{e-latt}=-2\sum_{{\vec k \, l=x,z} \atop {\xi=x,y,z} } \sqrt{\hbar \omega_{\vec k}} 
g_{\vec k \xi l} T_{-\vec kl} q_{\vec k \xi}\, .
\label{eq:JT2}
\end{eqnarray}
Here, 
$g_{\vec k \xi l}$ is defined as 
\begin{eqnarray}
g_{\vec k \xi l}={1 \over 2} {g_{JT} \over \sqrt{K}}(1-e^{-i k_{\xi}a}) C_{l \xi}, 
\end{eqnarray} 
with 
\begin{eqnarray}
C_{l \xi}=\left(
\begin{array}{ccc}
 {1 \over \sqrt{2}}, & -{1 \over \sqrt{2}}, & 0 \\
-{1 \over \sqrt{6}}, & -{1 \over \sqrt{6}}, &{2 \over \sqrt{6}}
\end{array}\right)_{l \xi}. 
\end{eqnarray}
Then, by using the canonical transformation, 
the linear couplings between $e_g$ electrons and lattice distortion are eliminated as 
\begin{eqnarray}
{\cal H}_{e-latt}+{\cal H}_{latt} 
&=& -{g_{JT}^2 \over K} 
\sum_{\vec k l l'} \widetilde A_{\vec k l l'}T_{-\vec k l}T_{\vec k l'} \nonumber \\
&&+\sum_{i \xi} {\hbar \omega_{\vec k} \over 2} 
\left( \widetilde p_{\vec k \xi}^2 + \widetilde q_{\vec k \xi}^2 \right) ,
\label{eq:Hoo}
\end{eqnarray}
where 
\begin{eqnarray}
\widetilde A_{\vec k l l'}={1 \over 2}\left(
\begin{array}{cc}
 2-c_x -c_y,                   & {1 \over \sqrt{3}}(c_x -c_y) \\
 {1 \over \sqrt{3}}(c_x -c_y), & {1 \over 3}(6-c_x -c_y -4c_z)
\end{array}\right)_{l l'} ,
\end{eqnarray}
and $c_{\rho}=\cos k_{\rho}a$. 
$\widetilde q_{\vec k \xi}$ is the new phonon coordinate given by 
\begin{eqnarray}
\widetilde q_{\vec k \xi}=q_{\vec k \xi} 
-{2 \over \sqrt{\hbar \omega_{\vec k}}}\sum_l g_{\vec k \xi l}^*T_{-\vec k l}\, , 
\end{eqnarray}
and 
$\widetilde p_{k \xi}$ is the canonical conjugate momentum for $\widetilde q_{k \xi}$. 
The first and second terms of the right hand side in Eq.~(\ref{eq:Hoo}) are denoted by 
$\widetilde {\cal H}_{o-o}$ and $\widetilde {\cal H}_{latt}$, respectively. 
Here, we neglect the noncommutability between 
${\cal H}$ (Eq.~(\ref{eq:Htot})) and $\widetilde q_{k \xi}$. 
$\widetilde {\cal H}_{o-o}$ includes the self-interaction of orbital, 
which does not contribute to the orbital order-disorder transition. 
Therefore, by subtracting this term, we obtain the following form: 
\begin{eqnarray}
{\cal H}_{o-o}=-{g_{JT}^2 \over K}\sum_{\vec k l l'} 
A_{\vec k l l'}T_{-\vec k l}T_{\vec k l'}\,,
\label{eq:JT4}
\end{eqnarray}
with 
\begin{eqnarray}
A_{k l l'}={1 \over 2}\left(
\begin{array}{cc}
-c_x -c_y,                    & {1 \over \sqrt{3}}(c_x -c_y) \\
{1 \over \sqrt{3}}(c_x -c_y), & {1 \over 3}(-c_x -c_y -4 c_z)
\end{array}\right)_{l l'}. 
\end{eqnarray}
Then, we obtain the effective Hamiltonian for the spin, orbital and lattice degrees of freedom 
in LaMnO$_3$ given by 
\begin{eqnarray}
{\cal H}_{eff}= {\cal H}_{e} +\widetilde{\cal H}_{latt} + {\cal H}_{o-o} + {\cal H}_{e-str} 
+ {\cal H}_{str} + {\cal H}_{hiJT}. 
\label{eq:Hmf}
\end{eqnarray}
\section{Mean Field Approximation}
In order to calculate the orbital ordering temperature from Eq.~(\ref{eq:Hmf}), 
we introduce the mean field approximation at finite temperatures. 
It is experimentally confirmed that the orbital order-disorder transition in LaMnO$_3$ 
is of the first order but is close to the second order transition; 
a discontinuity of the orbital order parameter at $T_{OO}$ is negligible.\cite{murakami} 
Therefore, we expect that the higher-order JT coupling, 
which brings about the first order phase transition, 
is much smaller than the linear JT coupling. 
Thus, we neglect ${\cal H}_{hiJT}$ in Eq.~(\ref{eq:Hmf}) and calculate $T_{OO}$. 
We will consider ${\cal H}_{hiJT}$ in the calculation of 
the elastic constant presented in Sec. V. 
\par
The $A$-type AF spin and $C$-type orbital orderings 
are observed in LaMnO$_3$.\cite{murakami,wollan,matsumoto} 
Two sublattices for the orbital (spin) ordering are denoted by $A$ and $B$ ($a$ and $b$) 
and the following mean fields are introduced: 
$\langle S_{a,b \, z} \rangle$, $\langle S^t_{a,b \, z} \rangle$, 
$\langle T_{A,B \, x} \rangle$ and $\langle T_{A,B \, z} \rangle$. 
The free energy of the system is obtained in the mean field approximation as follows: 
\begin{eqnarray}
{\cal F}_0&=&{Vc_0 \over 2}(u_x^2+u_z^2) \nonumber \\
&&-{N \over 2} \Bigl\{ 6J_{xy}^t \langle T_{Ax} \rangle \langle T_{Bx} \rangle 
+ 2J_{xy}^t \langle T_{Az} \rangle \langle T_{Bz} \rangle  \nonumber \\
&&\hspace{2.5em} + 2J_{z}^t \bigl( \langle T_{Az} \rangle^2 + \langle T_{Bz} \rangle^2 \bigr) \nonumber \\
&&\hspace{2.5em}+ J_2 \bigl( 2\langle S_{az} \rangle \langle S_{bz} \rangle - 
\langle S_{az} \rangle^2-\langle S_{bz} \rangle^2 \bigr) \nonumber \\
&&\hspace{3.5em}\times\bigl( \langle T_{Az} \rangle+\langle T_{Bz} \rangle \bigr)
\Bigr\} \nonumber \\
&& +{N \over 2} (J_1-3J_2) \bigl( \langle S_{az} \rangle^2+\langle S_{bz} \rangle^2  
+\langle S_{az} \rangle \langle S_{bz} \rangle \bigr) \nonumber \\
&&-NJ_{AF} \bigl( \langle S_{a z}^t \rangle^2 + \langle S_{a z}^t \rangle \langle S_{b z}^t \rangle 
+\langle S_{a z}^t \rangle^2 \bigr)
\nonumber \\
&&-{N \over 2 \beta}(\ln z_a^s + \ln z_b^s + \ln z_A^t + \ln z_B^t), 
\label{eq:Free0}
\end{eqnarray}
where $\beta=1/T$. 
$z_{a(b)}^s$ and $z_{A(B)}^t$ in 
the last four terms represent the partition functions of spin and orbital given by 
$z_{a(b)}^s={\rm Tr} \exp(-\beta {\cal H}_{a(b)}^s)$ and  
$z_{A(B)}^t={\rm Tr} \exp(-\beta {\cal H}_{A(B)}^t)$, respectively. 
${\cal H}_{a(b)}^s$ is the mean field Hamiltonian describing the spin state in sublattice $a$ $(b)$ as 
\begin{eqnarray}
{\cal H}_{a(b)}^s=
4 J_{xy}^s \langle \widetilde S_{a(b) \, z} \rangle \widetilde S_{a(b) \, z} +
2 J_{z}^s  \langle \widetilde S_{b(a) \, z} \rangle \widetilde S_{a(b) \, z}, 
\label{eq:Has}
\end{eqnarray}
where $J_{xy}^s$ and $J_{z}^s$ represent the effective exchange interaction between NN spins 
in the $xy$ plane and along the $z$ direction, respectively. 
These are explicitly given by 
\begin{eqnarray}
J_{xy}^s=&-&{1 \over 32}(J_1-3J_2) \nonumber \\
&+& {1 \over 32} (J_1+J_2) 
(3 \langle T_{Ax} \rangle \langle T_{Bx} \rangle + \langle T_{Az} \rangle \langle T_{Bz} \rangle) \nonumber \\
&-& {1 \over 16} J_2 (\langle T_{Az} \rangle+\langle T_{Bz} \rangle) 
+ {9 \over 16}J_{AF} , 
\label{eq:Jxy}
\end{eqnarray}
and
\begin{eqnarray}
J_{z}^s=&-&{1 \over 32}(J_1-3J_2) 
+ {1 \over 8} (J_1+J_2) \langle T_{Az} \rangle^2  \nonumber \\
&+& {1 \over 16} J_2 (\langle T_{Az} \rangle+\langle T_{Bz} \rangle) 
+ {9 \over 16}J_{AF} . 
\label{eq:Jz}
\end{eqnarray}
In Eq.~(\ref{eq:Has}), we introduce a spin operator $\vec {\widetilde S_i}$ with $\widetilde S=2$, 
rewrite $\vec S_i$ and $\vec S^t_i$ as 
$\vec S_i = {1 \over 4}\vec {\widetilde S_i}$ and $\vec S^t_i={3 \over 4}\vec {\widetilde S_i}$, 
respectively, and eliminate the largest energy parameter $J_H$ in Eq.~(\ref{eq:Hmf}). 
Due to the $A$-AF spin structure, the relation 
$\langle \widetilde S_{a z} \rangle =- \langle \widetilde S_{b z} \rangle$ is satisfied. 
${\cal H}_{A(B)}^t$ is the mean field Hamiltonian describing the orbital state in sublattice $A$ $(B)$ as 
\begin{eqnarray}
{\cal H}_{A(B)}^t &=&
6 J_{xy}^t \langle T_{B(A)x} \rangle T_{A(B)x} + 2 \tilde J T_{A(B)z} \nonumber \\
&&+\bigl( 2J_{xy}^t \langle T_{B(A)z} \rangle + 4J_{z}^t \langle T_{A(B)z} \rangle \bigr) T_{A(B)z} 
\nonumber \\
&&-2g_0\sqrt{Vc_0 \over N} (u_x T_{A(B)x} +u_z T_{A(B)z}) ,
\label{eq:Hat}
\end{eqnarray}
where $J_{xy(z)}^t$ and $\tilde J$ are 
\begin{eqnarray}
J_{xy(z)}^t={1 \over 4}(3 J_1 - J_2)
+(J_1 + J_2)
\langle S_{i z} \rangle \langle S_{i+\hat x(i+\hat z) z} \rangle 
\end{eqnarray}
and
\begin{eqnarray}
\tilde J = 2 J_2 \bigl( \langle S_{a z} \rangle^2 
- \langle S_{a z} \rangle\langle S_{b z} \rangle\bigr), 
\end{eqnarray}
respectively. 
For the observed $C$-type orbital ordered state, we have the following conditions: 
$\langle T_{Az}\rangle=\langle T_{Bz}\rangle$ and $\langle T_{Ax}\rangle=-\langle T_{Bx}\rangle$. 
\par
By minimizing the free energy ${\cal F}_0$ with respect to $\langle T_{A l} \rangle$ for $l=x, z$ 
and $\langle \widetilde S_{a z} \rangle$, 
the following self-consistent equations are obtained: 
\begin{eqnarray}
\langle T_{A l} \rangle &=& {\rm Tr} \Bigl\{ T_{A l} 
\exp\bigl(-\beta {\cal H}_A^t\bigr) \Bigr\}/z_A^t\, ,
\label{eq:t} \\
\langle \widetilde S_{a z} \rangle &=& 
{\rm Tr} \Bigl\{ \widetilde S_{a z} \exp\bigl(-\beta {\cal H}_a^s\bigr) \Bigr\}/z_a^s\, .
\label{eq:s}
\end{eqnarray}
Equations~(\ref{eq:t}) and (\ref{eq:s}) are numerically solved under the conditions of
$u_z=2g_0\sqrt{N/Vc_0}\langle T_{Az} \rangle$ and $u_x=0$
which are derived from 
$\partial{\cal F}/ \partial u_x=0$ and $\partial{\cal F}/ \partial u_x=0$, respectively. 
\section{Transition temperatures and spin wave dispersion}
Among several parameters in the Hamiltonian Eqs.~(\ref{eq:Htot}) and (\ref{eq:Hmf}), 
values of $J_1, J_2$ and $g_{JT}$ are determined by calculating 
the spin and orbital ordering temperatures and the spin wave dispersion relation. 
The other parameters are chosen to be 
$J_{AF}=1$, $a^2 K=17 \times 10^4$ and $a^3 c_0=2 \times 10^4$~meV, 
which are derived from 
the N{\'e}el temperature in CaMnO$_3,$\cite{wollan} 
the phonon frequency determined by the infrared absorption spectra\cite{arima} 
and the elastic constant.\cite{hazama} 
The lattice constant $a$ and the static JT distortion $Q (= \sqrt{Q_{i z}^2 +Q_{i x}^2})$ 
are chosen to be $a=4$~\AA \, and $Q=0.3$~\AA, respectively.\cite{mitchell} 
\par
Firstly, we calculate the transition temperatures. 
The orbital ordering temperature in the mean field approximation $T_{OO}^{MF}$ is given by 
\begin{eqnarray}
T_{OO}^{MF}={1 \over 2} \biggl\{ {3 \over 4} (3J_1-J_2) + {g_{JT}^2 \over K} \biggr\}, 
\label{eq:TOO}
\end{eqnarray}
and the N{\'e}el temperature for the $A$-AF ordering $T_N^{MF}$ is given by 
the solution of the following equation: 
\begin{eqnarray}
T_N^{MF} = -8 J_{xy}^s  +4 J_{z}^s, 
\label{eq:TN}
\end{eqnarray}
where $J_{xy}^s$ and $J_z^s$ defined in Eqs.~(\ref{eq:Jxy}) and (\ref{eq:Jz}) are the functions of 
$T_N^{MF}$. 
By fitting $T_{OO}^{MF}$ and $T_N^{MF}$ to the experimental transition temperatures 
$T_{OO}=780$~K (Ref.~\onlinecite{murakami}) and $T_N=140$~K 
(Refs.~\onlinecite{wollan} and \onlinecite{matsumoto}), 
respectively, 
$J_2$ and $g_{JT} Q$ are calculated as functions of $J_1$. 
In general, the mean field approximation tends to overestimate the transition temperature. 
Therefore, we also estimate the parameter values of $J_1$, $J_2$ and $g_{JT} Q$ 
by considering the correction of the mean field transition temperatures. 
For the N{\' e}el temperature, we revise $T_N^{MF}$ as 
$b T_N^{MF}$ with $b=0.63$, 
which is the ratio between $T_N$ for 
the $S=1/2$ AF Heisenberg model obtained in the high-temperature expansion and that 
in the mean field approximation.\cite{jou}
As for the orbital ordering temperature, we revise as $a T_{OO}^{MF}$ with $a=0.75$, 
which is obtained by the calculation of $T_N$ for 
the $S=1/2$ AF Ising model.\cite{jou}
This is because the orbital part of the Hamiltonian (Eq.~(\ref{eq:Hj})) has 
a discontinuous symmetry, although this symmetry is higher than that of the Ising model. 
\par
The spin stiffness constant provides another conditions for $J_1, J_2$ and $g_{JT}$.  
Although the JT distortion does not directly couple with the spin degree of freedom, 
$g_{JT}$ modifies the orbital state and affects the SE interaction between NN spins. 
We calculate the spin wave dispersion at $T=0$ and compare it with the experimental one. 
Here, the orbital and lattice degrees of freedom are assumed to be frozen, 
since the energy scale of orbital excitations\cite{ishihara1,e.saitoh} and optical phonon 
are much larger than that of spin wave. 
Then, the relevant parts of the Hamiltonian in Eq.~(\ref{eq:Htot}) are given by 
\begin{eqnarray}
{\cal H}_{SW}= {\cal H}_{e} + {\cal H}_{e-latt}\, . 
\label{eq:Hsw}
\end{eqnarray}
The static distortion of a MnO$_6$ octahedron is written as 
$(Q_{i z}, Q_{i x}) = Q (\cos \theta_i^{JT}, \sin \theta_i^{JT})$
where $\theta_A^{JT}=-\theta_B^{JT}={2\pi \over 3}$. 
The orbital ordered state is determined in the mean field approximation.\cite{okamoto} 
By applying the Holstein-Primakoff transformation to the spin operators, 
the dispersion relation of the spin wave is calculated. 
Experimentally, the spin wave in LaMnO$_3$ was measured by the neutron scattering experiments 
in Refs.~\onlinecite{hirota} and \onlinecite{moussa}. 
The authors in these papers analyzed the experimental data by using the Heisenberg model with 
the NN SE interactions. 
They obtained magnitude of the interaction 
in the $xy$ plane ($J_{xy}^s$) and that in the $z$ axis ($J_z^s$), 
which are defined in Eqs.~(\ref{eq:Jxy}) and (\ref{eq:Jz}), 
as $2 J_{xy}^s=-3.34$~meV and $2 J_z^s=2.42$~meV. 
\par
%
%
%
\begin{figure}[h]
\epsfxsize=0.8\columnwidth
\centerline{\epsffile{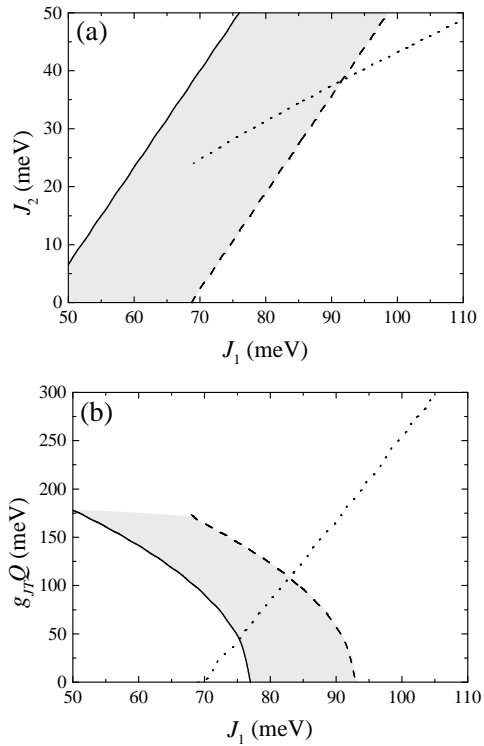}}
\caption{(a) $J_2$ and (b) $g_{JT}Q$ as functions of $J_1$. 
Solid and broken curves are obtained from 
the analyses of transition temperatures $T_{OO}$ and $T_N$. 
For broken curves, 
the correction from the mean field approximation is taken into account (see the text). 
The actual values of the parameters are expected to exist in the shaded regions. 
Dotted curves are obtained from the analysis of the spin stiffness constants. 
Other parameter values are chosen to be $J_{AF}=1$~meV, $a^2 K=17 \times 10^4$~meV, 
$B=0$~meV, $Q=0.3$~\AA \, and $a=4$~\AA. 
}
\label{fig:fig3}
\end{figure}
%
%
By fitting the theoretical results of the orbital ordering temperature, 
the N{\'e}el temperature for the $A$-AF ordering and the spin stiffness constant 
to the experimental values, we estimate $J_1, J_2$ and $g_{JT}$. 
$J_2$ and $g_{JT}Q$ are plotted as functions of $J_1$ in Fig.~\ref{fig:fig3}~(a) and (b), respectively. 
Solid and broken curves are obtained from the transition temperatures, $T_{OO}$ and $T_N$. 
The actual values of the parameters are expected to exist in the shaded regions. 
Dotted curves are obtained from the spin stiffness constants. 
The analyses for $T_{OO}$ and $T_N$ 
(solid and broken curves in Fig.~\ref{fig:fig3}~(a) and (b)) 
show that $J_2$ increases and $g_{JT}Q$ decreases with increasing $J_1$. 
This is because $(3J_1-J_2)$ and $g_{JT}^2/K$ contribute cooperatively to $T_{OO}$, 
as shown in Eq.~(\ref{eq:TOO}). 
On the other hand, the analyses for the spin stiffness constant 
(dotted curves in Fig.~\ref{fig:fig3}~(a) and (b)) show that 
both $J_2$ and $g_{JT}Q$ increase with increasing $J_1$, 
i.e., $J_1$ competes with both $J_2$ and $g_{JT}Q$. 
This is attributed to the facts that (1) $J_1$ and $J_2$ are the ferromagnetic and 
antiferromagnetic interactions, respectively, 
and (2) $g_{JT}Q$ favors the $(d_{3x^2-r^2}/d_{3y^2-r^2})$-type orbital ordered state 
where the ferromagnetic interaction in the $xy$ plane caused by $J_1$ is 
weaker than that without $g_{JT}Q$. 
The magnitudes of $J_1, J_2$ and $g_{JT}Q$ are obtained 
from the intersection points of solid (broken) curve and dotted curve in Fig.~\ref{fig:fig3}~(a) and (b).  
We obtain $J_1=75 \sim 85$~meV, $J_2=25 \sim 40$~meV and $E_{JT}=g_{JT}Q= 50\sim100$~meV. 
It is stressed that the value of $E_{JT}$ is much smaller than 
that in the literature, i.e., $E_{JT}>1$~eV.\cite{millis,jung,machida,quijada,allen,coey,ahn,satpathy,popovic,hozoi} 
In the case that $a^2K=17 \times 10^4$~meV, 
the coupling constant of the inter-site orbital interaction through the exchange of phonons 
$g_{JT}^2/K$ is about $3 \sim 10$~meV which is much smaller than $J_1$. 
We conclude that the small $E_{JT}$ comes from the strong Coulomb interaction 
and the orbital ordering in LaMnO$_3$ is dominated by the interaction through 
the virtual exchange of electrons. 
If $J_1$ and $J_2$ are neglected to estimate $T_{OO}$, 
we obtain $E_{JT}$ to be 
$400 \sim 700$~meV for $a^2K = 17 \times 10^4$~meV\cite{arima} 
and $600 \sim 1100$~meV for $a^2K = 40 \times 10^4$~meV.\cite{millis} 
The latter value of $E_{JT}$ is of the same order of magnitude given in Ref.~\onlinecite{millis}. 
\section{Elastic Constants}
The elastic constants provide information of the higher-order JT coupling,\cite{kataoka,kataoka2} 
although the coupling constant $B$ is supposed to be smaller than that in the linear JT coupling. 
In this section, we examine the elastic constants taking into account 
the electron-electron and electron-lattice interactions and the higher-order JT coupling. 
\par
We start with the model Hamiltonian in Eq.~(\ref{eq:Hmf}). 
The elastic constants are the coefficients of the $\delta u_l^2$ terms in the free energy, 
where $\delta u_l$ is the deviation of strain $u_l$ from that in the thermal equilibrium. 
Here, the deviations of $T_{A l}$ and $T_{B l}$ from the thermal equilibrium, 
$\delta T_{A l}$ and $\delta T_{B l}$, are also introduced. 
$\delta T_{A l}$ and $\delta T_{B l}$ are induced by an external strain $\delta u_{l'}$ 
and the relations between them are derived later. 
Now, the free energy is expanded up to the second order of $\delta T_{A l}$, $\delta T_{B l}$ and 
$\delta u_l$ as 
\begin{eqnarray}
{\cal F}&=&{\widetilde{\cal F}}_0+{Vc_0 \over 2}\bigl( \delta u_x^2 + \delta u_z^2 \bigr) \nonumber \\
&&-{N \over 2} \Bigl[ 6J_{xy}^t \delta T_{Ax} \delta T_{Bx} + 2J_{xy}^t \delta T_{Az} \delta T_{Bz} 
\nonumber \\
&& \hspace{3em}
+2J_z^t \bigl\{(\delta T_{Az}+\delta T_{Bz})^2 - (\delta T_{Az}-\delta T_{Bz})^2 \bigr\} \Bigr] 
\nonumber \\
&&-{\beta N \over 4} \biggl\{ 
F_{xx} \Bigl( {C_x^A}^2 + {C_x^B}^2 \Bigr) + F_{zz} \Bigl( {C_x^A}^2 + {C_x^B}^2 \Bigr) \nonumber \\
&& \hspace{3.5em}
+2 F_{xz} 
\Bigl( C_x^A C_z^A -C_x^B C_z^B \Bigr)
\biggr\}. 
\label{eq:FreeD}
\end{eqnarray}
Here, $\widetilde{\cal F}_0$ is given by Eq.~(\ref{eq:Free0}) where 
$z_{A(B)}^t$ is replaced by $z_{A(B)}^t={\rm Tr} \exp(-\beta \widetilde {\cal H}_{A(B)}^t)$, 
$\widetilde {\cal H}_{A(B)}^t$ being given by 
\begin{eqnarray} 
\widetilde {\cal H}_{A(B)}^t
&=&{\cal H}_{A(B)}^t \nonumber \\
&&-B{Vc_0 \over Ng_0} \Bigl\{ \bigl( Q_{A(B)z}^2-Q_{A(B)x}^2 \bigr)T_{A(B)z} \nonumber \\
&&\hspace{5em}-2Q_{A(B)z}Q_{A(B)x}T_{A(B)x} \Bigr\}. 
\label{eq:Hatnew}
\end{eqnarray}
The last term of this equation comes from the higher-order JT coupling. 
We assume that the equilibrium values of $Q_{Al}$ appearing in this term are given by 
$Q_{Az}=au_z$ and 
$Q_{Ax}=4{g_{JT} \over K} \langle T_{Ax} \rangle$ by considering the definition of 
the strain $u_z$ and the linear JT coupling in Eq.~(\ref{eq:Helatt}). 
$C_l^{\alpha}$'s are the coefficients of $T_{\alpha l}$ 
in ${\cal H}_{\alpha}^t$ (Eq.~(\ref{eq:Hat})) 
where $\langle T_{A l} \rangle$, $\langle T_{B l} \rangle$ and 
$u_l$ are replaced by $\langle T_{Al} \rangle + \delta T_{Al}$, 
$\langle T_{Bl} \rangle + \delta T_{Bl}$ and $u_l + \delta u_l$, respectively. 
Their explicit forms are given by 
\begin{eqnarray}
C_x^A&=&6J_{xy}^t \delta T_{Bx}-2g_0 \sqrt{Vc_0 \over N} \delta u_x , \\
C_z^A&=&2J_{xy}^t \delta T_{Bz} + 4J_z^t \delta T_{Az} -2g_0 \sqrt{Vc_0 \over N} \delta u_z , \\
C_x^B&=&6J_{xy}^t \delta T_{Ax}-2g_0 \sqrt{Vc_0 \over N} \delta u_x , \\
C_z^B&=&2J_{xy}^t \delta T_{Az} + 4J_z^t \delta T_{Bz} -2g_0 \sqrt{Vc_0 \over N} \delta u_z. 
\end{eqnarray}
Here, the term originating from the higher-order JT coupling is neglected.\cite{kataoka} 
$F_{ll'}$ in Eq.~(\ref{eq:FreeD}) represents the self-correlation function of the orbital given by 
\begin{eqnarray}
F_{ll'}=\langle T_{Al} \rangle \langle T_{Al'} \rangle -K_{ll'} 
\end{eqnarray}
with 
\begin{eqnarray}
K_{l l'}
&=& {1 \over z_A^t} 
\Biggl[ \sum_{m m'} e^{-\beta \varepsilon_m} 
\delta_{\varepsilon_m,\varepsilon_{m'}}
-{2 \over \beta} \hspace{-1em} \sum_{{m>m'} \atop 
{(\varepsilon_m \ne \varepsilon_{m'})}} 
\hspace{-1em}
{e^{-\beta \varepsilon_m}-e^{-\beta \varepsilon_{m'}} \over 
\varepsilon_m - \varepsilon_{m'}} 
\Biggr] \nonumber \\
&& \hspace{2em} \times (T_{A l})_{mm'} (T_{A l'})_{m'm} , 
\label{eq:Kmm}
\end{eqnarray}
where $\varepsilon_m$ represents the $m$ th eigenvalue of $\widetilde {\cal H}_A^t$. 
To derive Eq.~(\ref{eq:FreeD}), 
the conditions $\langle T_{Ax} \rangle=-\langle T_{Bx} \rangle$, 
$\langle T_{Az} \rangle=\langle T_{Bz} \rangle$, $Q_{Az}=Q_{Bz}$ and $Q_{Ax}=-Q_{Bx}$ are used. 
By using the condition $\partial {\cal F}/\partial \delta T_{\alpha l}=0$, 
the following relations between $\delta u_l$'s and $\delta T_{\alpha l'}$'s are obtained: 
\begin{eqnarray}
\delta T_{Ax} + \delta T_{Bx}
&=&-{4 \beta g_0 \over D_x} \sqrt{Vc_0 \over N} \delta u_x \nonumber \\
&& \hspace{-6em} \times 
\biggl[ F_{xx} \Bigl\{ 1- \bigl( 4J_z^t-2J_{xy}^t \bigr) \beta F_{zz} \Bigr\} 
+ \bigl( 4J_z^t-2J_{xy}^t \bigr) \beta F_{xz}^2 \biggr], \nonumber \\
\label{eq:delta1} \\
\delta T_{Az} - \delta T_{Bz}
&=&-{4 \beta g_0 \over D_x}\sqrt{Vc_0 \over N} \delta u_x F_{xz} 
\label{eq:delta2}, \\
\delta T_{Az} + \delta T_{Bz}
&=&-{4 \beta g_0 \over D_z}\sqrt{Vc_0 \over N} \delta u_z \nonumber \\
&&\times 
\Bigl\{ F_{zz} \bigl( 1- 6J_{xy}^t \beta F_{xx} \bigr) + 6J_{xy}^t \beta F_{xz}^2 \Bigr\} 
\label{eq:delta3}, \\
\delta T_{Ax} - \delta T_{Bx}
&=&-{4 \beta g_0 \over D_z} \sqrt{Vc_0 \over N} \delta u_z F_{xz} , 
\label{eq:delta4} 
\end{eqnarray} 
with 
\begin{eqnarray}
D_x&=&
\bigl( 1- 6J_{xy}^t \beta F_{xx} \bigr) 
\Bigl\{ 1- \bigl( 4J_z^t-2J_{xy}^t \bigr) \beta F_{zz} \Bigr\} \nonumber \\ 
&&- 6 J_{xy}^t \bigl( 4J_z^t-2J_{xy}^t \bigr) \beta^2 F_{xz}^2 \, , 
\end{eqnarray}
and
\begin{eqnarray}
D_z&=&
\bigl( 1+ 6J_{xy}^t \beta F_{xx} \bigr) 
\Bigl\{ 1- \bigl( 4J_z^t+2J_{xy}^t \bigr) \beta F_{zz} \Bigr\} \nonumber \\ 
&&+ 6J_{xy}^t \bigl( 4J_z^t+2J_{xy}^t \bigr) \beta^2 F_{xz}^2 \, . 
\end{eqnarray}
By using Eqs.~(\ref{eq:delta1})-(\ref{eq:delta4}), 
the deviation of the free energy from the equilibrium value is given by 
\begin{eqnarray}
{\cal F}-\widetilde {\cal F}_0 =
{1 \over 2} \bigl( c_x(T) \delta u_x^2+c_z(T) \delta u_z^2 \bigr) .
\label{eq:FreeDD}
\end{eqnarray}
Note that the term proportional to $\delta u_x \delta u_z$ is absent because of the tetragonal symmetry. 
$c_x(T)$ and $c_z(T)$ are the temperature dependent elastic constants for 
the $u_x$ and $u_z$ modes, respectively. 
Their explicit forms are given by 
\begin{eqnarray}
&&c_x(T)\nonumber \\
&&={c_0 \over D_x}
\biggl[ \Bigl\{ 1- \bigl( 6J_{xy}^t - 4g_0^2 \bigr) \beta F_{xx} \Bigr\}
\Bigl\{ 1- \bigl( 4J_z^t-2J_{xy}^t \bigr) \beta F_{zz} \Bigr\} \nonumber \\
&& \hspace{3em} 
-\bigl( 6 J_{xy}^t - 4g_0^2 \bigr) \bigl( 4J_z^t-2J_{xy}^t \bigr) \beta^2F_{xz}^2 \biggr], 
\end{eqnarray}
and
\begin{eqnarray}
&&c_z(T)\nonumber \\
&&={c_0 \over D_z}
\biggl[ \bigl( 1+ 6J_{xy}^t \beta F_{xx} \bigr)
\Bigl\{ 1- \bigl( 4J_z^t+2J_{xy}^t -4g_0^2 \bigr) \beta F_{zz} \Bigr\} \nonumber \\
&& \hspace{3em} 
+ 6 J_{xy}^t \bigl( 4J_z^t+2J_{xy}^t -4g_0^2 \bigr) \beta^2F_{xz}^2 \biggr].
\end{eqnarray}
For $T > T_{OO}$, $c_x(T)$ and $c_z(T)$ are simplified as 
\begin{eqnarray}
c_x(T)=c_z(T)=c_0\frac{T+{3 \over 8}(3J_1-J_2)+{1 \over 2}{g_{JT}^2 \over K}-g_0^2}
{T+{3 \over 8}(3J_1-J_2)+{1 \over 2}{g_{JT}^2 \over K}}.
\end{eqnarray}
\par
%
%
%
\begin{figure}
\epsfxsize=0.8\columnwidth
\centerline{\epsffile{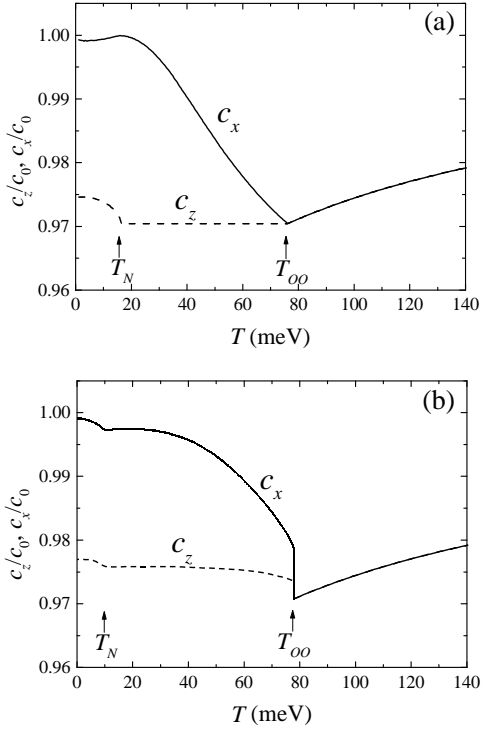}}
\caption{The elastic constants as functions of $T$. 
Solid and broken curves denote $c_x$ and $c_z$, respectively. 
Parameter values are chosen to be 
$J_1=75$~meV, $J_2=25$~meV, $J_{AF}=1$~meV, $a g_{JT}=6 \times 10^2$~meV, $a^2 K=17 \times 10^4$~meV 
and $a^3 c_0=2 \times 10^4$~meV. 
The value of $B$ is chosen as $B=0$~meV in (a) and $B=50$~meV in (b). 
}
\label{fig:fig4}
\end{figure}
%
%
%
%
\begin{figure}
\epsfxsize=0.8\columnwidth
\centerline{\epsffile{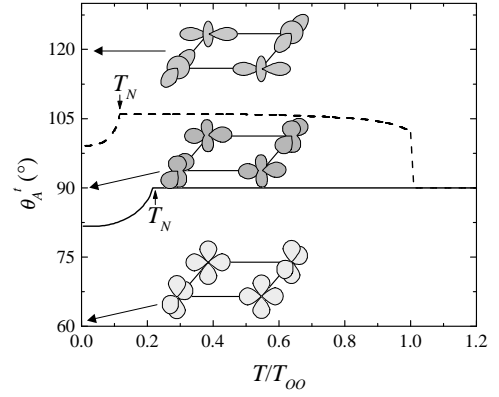}}
\caption{Temperature dependence of the orbital state. 
$\theta_A^t$ is defined by $\theta_A^t = \tan^{-1} ( \langle T_{Ax} \rangle / \langle T_{Az} \rangle )$. 
Solid and broken curves are the results for $B=0$~meV and $B=50$~meV, respectively. 
Other parameter values are the same as those in Fig.~\ref{fig:fig4}. 
The insets show the schematic orbital structures in the $xy$ plane. 
}
\label{fig:fig5}
\end{figure}
%
%

The elastic constants are numerically calculated and presented as functions of $T$ 
in Fig.~\ref{fig:fig4}. 
Parameter values are chosen to be 
$J_1=75$~meV, $J_2=25$~meV, $J_{AF}=1$~meV, $a g_{JT}=6 \times 10^2$~meV, $a^2 K=17 \times 10^4$~meV 
and $a^3 c_0=2 \times 10^4$~meV. 
The higher-order JT coupling $B$ is chosen to be 
$B=0$~meV in Fig.~\ref{fig:fig4}~(a) and $B=50$~meV in Fig.~\ref{fig:fig4}~(b). 
When $B$ is comparable to $J_1$, 
the discontinuous change in the elastic constants is found at $T_{OO}$. 
We also obtain a change in $c_x$ at $T_N$. 
This change reflects the change of the orbital state. 
As shown in Fig.~\ref{fig:fig5}, when $B$ is much smaller than $J_1$ (solid curve), 
the orbital ordered state of 
the ${1 \over \sqrt{2}}(d_{3z^2-r^2}+d_{x^2-y^2} / d_{3z^2-r^2}-d_{x^2-y^2})$-type above $T_N$ 
changes to the state below $T_N$ where the component of the ($d_{y^2-z^2}/ d_{z^2-x^2}$)-type 
increases. 
On the other hand, when $B$ is comparable to $J_1$ (broken curve), 
the component of the ${1 \over \sqrt{2}}(d_{3z^2-r^2}+d_{x^2-y^2} / d_{3z^2-r^2}-d_{x^2-y^2})$-type, 
state increases below $T_N$. 
These changes of the orbital states originate from the spin-orbital coupling in ${\cal H}_J$ 
in Eq.~(\ref{eq:Hj}). 
The changes of the elastic constants are understood as follows: 
In the ${1 \over \sqrt{2}}(d_{3z^2-r^2}+d_{x^2-y^2} / d_{3z^2-r^2}-d_{x^2-y^2})$-type 
orbital ordered state, 
$\langle T_{Ax} \rangle$ is almost saturated as $|\langle T_{Ax} \rangle|= 1/2$ 
and $\langle T_{Az} \rangle= 0$. 
Thus, the external strain $\delta u_x$ does not induce $\delta T_{Ax}$ 
and $c_x$ is saturated. 
The deviation of the orbital state from 
the ${1 \over \sqrt{2}}(d_{3z^2-r^2}+d_{x^2-y^2} / d_{3z^2-r^2}-d_{x^2-y^2})$-type one 
causes the decrease of $c_x$ as shown in Fig.~\ref{fig:fig4}. 
This fact reminds us of the inverse of the parallel spin susceptibility $1/\chi_{\parallel}$ 
diverging with decreasing $T$ in an antiferromanget. 
We propose that the characteristic change of the elastic constant at $T_N$ 
may be used to estimate the coupling constant $B$. 
\section{Summary and discussion}
We have examined the orbital ordering in LaMnO$_3$ and 
the magnitudes of the interactions of the $e_g$ orbitals between NN sites 
caused by the virtual exchange of electrons and phonons. 
By calculating the orbital and spin ordering temperatures 
and the spin wave dispersion and comparing them with the experimental results, 
we obtained $J_1=75 \sim 85$~meV, $J_2=25 \sim 40$~meV and $g_{JT}Q=E_{JT}= 50\sim100$~meV. 
$E_{JT}$ is much smaller than that in 
the literature\cite{millis,jung,machida,quijada,allen,coey,ahn,satpathy,popovic,hozoi} 
which was estimated by neglecting the electron-electron interaction. 
The present results indicate that the orbital ordering in LaMnO$_3$ is mainly caused by 
virtual exchange of electrons under the strong Coulomb interactions.\cite{benedetti} 
We calculate the temperature dependence of the elastic constants by taking into account 
both the electron-electron and electron-lattice interactions. 
It is predicted that the elastic constants show the characteristic change at $T_N$ 
which depends on the magnitude of the higher-order JT coupling; 
when the coupling constant $B$ is comparable to $J_1$ (much smaller than $J_1$), 
$c_x$ increases (decreases) with decreasing $T$. 
Through the detailed comparison between theory and experiment, 
the value of $B$ may be determined. 
\par
The present results support the recent report on the observation of 
the collective orbital excitation termed orbital wave in LaMnO$_3$ by the Raman scattering experiments. 
Saitoh {\it et al}. have observed three peak structures around 120-160~meV in 
the Raman spectra.\cite{e.saitoh} 
These peaks can be explained by neither the two-phonon excitations nor 
the magnetic excitations. 
The theoretical results of the Raman spectra from 
the orbital wave agree with the polarization dependence of the spectra 
and their relative intensities in experiment.\cite{okamotoR} 
Since the characteristic energy of the orbital excitation is much higher 
than that of the lattice vibration, 
we introduce the adiabatic approximation for the lattice degree of freedom 
in the calculation of the orbital wave. 
Then, the energy of orbital wave is approximately given by  
$\omega_{orb}=
\sqrt{ (3J_1+{\sqrt{3} \over 2}E_{JT}) (J_1+J_2+{\sqrt{3} \over 2}E_{JT})}$. 
When we adopt the parameter values obtained in the present analyses 
$J_1=75$~meV, $J_2=25$~meV and $E_{JT}=50$~meV, we obtain $\omega_{orb}=187$~meV. 
These numerical values are consistent with the observed energies of 
the Raman shifts from the orbital excitations. 
\acknowledgments
The authors would like to thank T.~A.~Kaplan, T.~Goto, H.~Hazama, N.~Nagaosa, Y.~Tokura and E.~Saitoh 
for their valuable discussions. 
This work was supported by Grant-in-Aid for Scientific Research Priority Area from 
the Ministry of Education, Science, Sports, Culture and Technology of Japan, CREST Japan 
and Science and Technology Special Coordination Fund for Promoting Science and Technology. 
Part of the numerical calculation was performed in 
the supercomputing facilities in IMR, Tohoku University. 
S.~M. acknowledges support of the Humboldt Foundation. 
%
%
%


\begin{references}
%
%
\bibitem{science} Y.~Tokura and N.~Nagaosa, 
Science {\bf 288}, 462 (2000).
%
\bibitem{chahara}
K.~Chahara, T.~Ohno, M.~Kasai, Y.~Kanke, and Y.~Kozono, 
Appl.\ Phys.\ Lett. {\bf 62}, 780 (1993).
%
\bibitem{helmolt}
R.~von Helmolt, J.~Wecker, B.~Holzapfel, L.~Schultz, and K.~Samwer, 
Phys.\ Rev.\ Lett. {\bf 71}, 2331 (1993). 
%
\bibitem{tokura}
Y.~Tokura, A.~Urushibara, Y.~Moritomo, T.~Arima, A.~Asamitsu, G.~Kido, and N.~Furukawa, 
J.\ Phys.\ Soc.\ Jpn. {\bf 63}, 3931 (1994).
%
\bibitem{murakami}
Y.~Murakami, J.~P.~Hill, D.~Gibbs, M.~Blume,  I.~Koyama, M.~Tanaka, H.~Kawata, T.~Arima, 
Y.~Tokura, K.~Hirota, and Y.~Endoh, 
Phys.\ Rev.\ Lett. {\bf 81}, 582 (1998). 
%
\bibitem{wollan}
E.~O.~Wollan and W.~C.~Koehler, 
Phys.\ Rev. {\bf 100}, 545 (1955). 
%
\bibitem{matsumoto}
G. Matsumoto,  
J.\ Phys.\ Soc.\ Jpn. {\bf 29}, 606 (1970).
%
\bibitem{goodenough}
J. B. Goodenough, 
Phys.\ Rev. {\bf 100}, 564 (1955), and 
in {\it Progress in Solid State Chemistry}, edited 
by H. Reiss (Pergamon, London, 1971), Vol. 5.
%
\bibitem{kanamori} 
J.~Kanamori, 
J.\ Phys.\ Chem.\ Sol. {\bf 10}, 87 (1959).
%
\bibitem{kugel}
K.~I.~Kugel and D.~I.~Khomskii, 
Zh.\ {\'E}ksp. Teor. Fiz. {\bf 64}, 1429 (1973) [Sov. Phys. JETP {\bf 37}, 725 (1973)]. 
%
\bibitem{hirota} 
K.~Hirota, N.~Kaneko, A. Nishizawa, and Y. Endoh, 
J.\ Phys.\ Soc.\ Jpn. {\bf 65}, 3736 (1996).
%
\bibitem{ishihara1}
S.~Ishihara, J.~Inoue, and S.~Maekawa, 
Physica C {\bf 263}, 130 (1996), and Phys.\ Rev.\ B {\bf 55}, 8280 (1997).
%
\bibitem{rodriguez}
J.~Rodriguez-Carvajal, M.~Hennion, F.~Moussa, A.~H.~Moudden, L.~Pinsard, and A.~Revcolevschi
Phys.\ Rev.\ B {\bf 57}, R3189 (1998). 
%
\bibitem{maezono}
R.~Maezono, S.~Ishihara, and N.~Nagaosa, 
Phys.\ Rev.\ B {\bf 57}, R13993 (1998).
%
\bibitem{feinberg}
D.~Feinberg, P.~Germain, M.~Grilli, and G.~Seibold, 
Phys.\ Rev.\ B {\bf 57}, R5583 (1998). 
%
%
\bibitem{kanamori2} 
J.~Kanamori, 
J.\ Appl.\ Phys. {\bf 31}, 14S (1960).
%
\bibitem{kataoka}
M.~Kataoka and J.~Kanamori, 
J.\ Phys.\ Soc.\ Jpn. {\bf 32}, 113 (1972).
%
\bibitem{millis} 
A.~J.~Millis, 
Phys.\ Rev.\ B {\bf 53}, 8434 (1996). 
%
\bibitem{kataoka2}
M.~Kataoka, 
J.\ Phys.\ Soc.\ Jpn. {\bf 70}, 2353 (2001).
%
\bibitem{nagaosa}
N.~Nagaosa, S.~Murakami, H.~C.~Lee, 
Phys.\ Rev.\ B {\bf 57}, R6767 (1998). 
%
\bibitem{benedetti}
P.~Benedetti and R.~Zeyher, 
Phys.\ Rev.\ B {\bf 59}, 9923 (1999). 
%
\bibitem{bala}
J.~Ba{\l}a and A.~M.~Ole{\'s}, 
Phys.\ Rev.\ B {\bf 62}, R6085 (2000). 
%
\bibitem{jung}
J.~H.~Jung, K.~H.~Kim, T.~W.~Noh, E.~J.~Choi, and Jaejun Yu, 
Phys.\ Rev.\ B {\bf 57}, R11043 (1998). 
%
\bibitem{machida}
A.~Machida, Y.~Moritomo, and A.~Nakamura, 
Phys.\ Rev.\ B {\bf 58}, 12540 (1998). 
%
\bibitem{quijada}
M.~Quijada, J.~{\v C}erne, J.~R.~Simpson, H.~D.~Drew, K.~H.~Ahn, A.~J.~Millis, R.~Shreekala, R.~Ramesh, 
M.~Rajeswari, and T.~Venkatesan, 
Phys.\ Rev.\ B {\bf 58}, 16093 (1998). 
\bibitem{allen}
P.~B.~Allen and V.~Perebeinos, 
Phys.\ Rev.\ Lett. {\bf 83}, 4828 (1999).
%
\bibitem{coey}
J.~M.~Coey, M.~Viret, and S.~von Moln{\' a}r, 
Adv.\ Phys.\ {\bf 48}, 167 (1999). 
%
\bibitem{ahn}
K.~H.~Ahn and A.~J.~Millis, 
Phys.\ Rev.\ B {\bf 61}, 13545 (2000). 
%
\bibitem{satpathy}
S.~Satpathy, Z.~S.~Popovi{\'c}, and F.~R.~Vukajlovi{\'c}, 
J.\ Appl.\ Phys. {\bf 79}, 4555 (1996).
%
\bibitem{popovic}
Z.~Popovic and S.~Satpathy, 
Phys.\ Rev.\ Lett.\ {\bf  84},  1603 (2000).  
%
\bibitem{hozoi}
L.~Hozoi, A.~H.~de Vries, and R.~Broer, 
Phys.\ Rev.\ B {\bf 64}, 165104 (2001). 
%
\bibitem{t.saitoh}
T.~Saitoh, A.~E.~Bocquet, T.~Mizokawa, H.~Namatame, A.~Fujimori, M.~Abbate, Y.~Takeda, and M.~Takano, 
Phys.\ Rev.\ B {\bf  51},  13942 (1995).  
%
\bibitem{okamoto}
S.~Okamoto, S.~Ishihara, and S.~Maekawa, 
Phys.\ Rev.\ B {\bf 61}, 451 (2000), {\it ibid.} 14647 (2000). 
%
\bibitem{moussa} 
F.~Moussa, M.~Hennion, J.~Rodriguez-Carvajal, H.~Moudden, L.~Pinsard, and A.~Revcolevschi, 
Phys.\ Rev.\ B {\bf 54}, 15149 (1996).
%
\bibitem{arima}
T.~Arima, Y.~Tokura, and J.~B.~Torrance, 
Phys.\ Rev.\ B {\bf 48}, 17006 (1993); 
T.~Arima and Y.~Tokura, 
J.\ Phys.\ Soc.\ Jpn. {\bf 64}, 2488 (1995). 
%
\bibitem{hazama}
H.~Hazama, T.~Goto, Y.~Nemoto, Y.~Tomioka, A.~Asamitsu and Y.~Tokura, 
Phys.\ Rev.\ B {\bf 62}, 15012 (2000). 
%
\bibitem{mitchell}
J.~F.~Mitchell, D.~N.~Argyriou, C.~D.~Potter, D.~G.~Hinks, J.~D.~Jorgensen, and S.~D.~Bader,
Phys.\ Rev.\ B {\bf 54}, 6172 (1996). 
%
\bibitem{jou}
D.~C.~Jou and H.~H.~Chen, 
Phys.\ Lett. {\bf 45A}, 239 (1973).
%
\bibitem{e.saitoh}
E.~Saitoh, S.~Okamoto, K.~Takahashi, K.~Tobe, K.~Yamamoto, T.~Kimura, S.~Ishihara, S.~Maekawa, 
and Y.~Tokura, 
Nature {\bf 410}, 180 (2001). 
%
\bibitem{okamotoR}
S.~Okamoto, S.~Ishihara, and S.~Maekawa, 
(unpublished) cond-mat/0108032. 
%
\end{references}
\end{document}